\pdfoutput=1
\documentclass[conference]{IEEEtran}
\IEEEoverridecommandlockouts
\usepackage{cite}
\usepackage{amsmath,amssymb,amsfonts}
\usepackage{algorithmic}
\usepackage{graphicx}
\usepackage{textcomp}
\usepackage{xcolor}

\definecolor{mygreen}{RGB}{28,172,0}
\usepackage{listings}
\def\BibTeX{{\rm B\kern-.05em{\sc i\kern-.025em b}\kern-.08em
    T\kern-.1667em\lower.7ex\hbox{E}\kern-.125emX}}
\begin{document}

\title{Design and Implementation of Knowledge Base for Runtime Management of Software Defined Hardware\\
% \thanks{SDH Program, Defense Advanced Research Projects Agency. DDARING Project, Georgia Institute of Technology
% }
}

\author{\IEEEauthorblockN{Hongkuan Zhou$^\dagger$, Ajitesh Srivastava$^\dagger$, Rajgopal Kannan$^{\ddagger}$, Viktor Prasanna$^\dagger$}
\IEEEauthorblockA{\textit{$^\dagger$University of Southern California \ \ \ \ $^{\ddagger}$Army Research Lab-West} \\
Los Angeles, USA \\
\{hongkuaz, ajiteshs, rajgopak, prasanna\}@usc.edu} %rajgopal.kannan.civ@mail.mil
% \and
% \and
% \and
% \and
% \and
}

\maketitle

\begin{abstract}

Runtime-reconfigurable software coupled with reconfigurable hardware is highly desirable as a means towards maximizing runtime efficiency without compromising programmability. Compilers for such software systems are extremely difficult to design as they must leverage different types of hardware at runtime. To address the need for static and dynamic compiler optimization of workflows matched to dynamically reconfigurable hardware, we propose a novel design of the central component of a dynamic software compiler for software defined hardware. Our comprehensive design focuses not just on static knowledge but also on semi-supervised extraction of knowledge from program executions and developing their performance models. Specifically, our novel {\it dynamic and extensible  knowledge base} 1) continuously gathers knowledge during execution of workflows 2) identifies {\it optimal} implementations of workflows on  {\it optimal} (available) hardware configurations.
It plays a hub role in storing information from, and providing information to other components of the compiler, as well as the human analyst. Through a rich tripartite graph representation, the knowledge base captures and learns extensive information on decomposition and mapping of code steps to kernels and mapping of kernels to available hardware configurations. The knowledge base is implemented using the C++ Boost Library and is capable of quickly processing offline and online queries and updates. We show that our knowledge base can answer queries in $1ms$ regardless of the number of workflows it stores. To the best of our knowledge, this is the first design of a dynamic and extensible knowledge base to support compilation of high-level languages to leverage arbitrary reconfigurable platforms.

\end{abstract}


\section{Introduction}

The modern world is driven by information - the increasing availability of information  must be matched by large scale computational algorithms for processing this data.
With the emergence of diverse, heterogeneous system architectures, a major challenge for algorithm developers is balancing the tradeoff between algorithm runtime efficiency and ease of implementation on available hardware. 
Consider two extreme cases - general purpose CPUs versus application specific integrated circuits (ASICs). Developers can write in a concise high-level programming language to efficiently implement algorithms on CPUs. Thus the former offers very good programmability but (relatively) low compute efficiency.  Conversely, ASICs, being specialized, maximize runtime efficiency at the cost of poor programmability. Indeed, due to the low flexibility of ASICs, only high priority algorithms are considered for implementation. 
For other algorithms, one has to sacrifice  compute efficiency by implementing on alternative  hardware like field programmable gate arrays (FPGAs) or CPUs. These algorithms can run much slower than on ASICs.
This poses the question, {\it how to balance the tradeoff between programmability and run-time efficiency for reconfigurable hardware} which runs the spectrum between CPUs and ASICs.

% Taking the advantage of CPUs and ASICs, the software defined hardware is a middle layer of software which is in charge of configuring the hardware. The Software Defined Hardware (SDH) \cite{SDH} program aims to design a software define hardware that is optimized for a collection of data intensive workflows.

\subsection{SDH Program}
\label{sec:sdh}

% The goal of the SDH \cite{SDH} program is to build runtime-reconfigurable hardware and software that enables near ASIC performance without sacrificing programmability for data-intensive algorithms. Under the program, data-intensive algorithms are defined as machine learning and data science algorithms that process large volumes of data and are characterized by their usage of intense linear algebra, graph search operations, and their associated data-transformation operators. The SDH program aims to create hardware/software systems that allow data-intensive algorithms to run at near ASIC efficiency without the cost, development time, or single application limitations associated with ASICs. If successful, SDH will result in the ability to develop and run data-intensive, data-exploitation algorithms at very low cost, and, consequently, enable pervasive use of big-data solutions for a wide range of DoD applications including ISR, predictive logistics, decision support, and beyond.

To achieve both high programmability and compute efficiency, the key is to {\it build a system consisting of tightly coupled hardware and software}. The DARPA Software Defined Hardware (SDH) \cite{SDH} program aims to combine runtime-reconfigurable hardware with a dynamic (reconfigurable) software compiler that achieves comparable compute efficiency to ASICs but without the associated cost and time of development, or application field-limitations. The target algorithms of the DARPA SDH program are a selection of data-intensive workflows from the domains of image, video, audio, text, signal, and graph.
% These workflows, including machine learning and data science algorithms, should be able to develop and run at a very low cost in the software-defined hardware which would enable pervasive use of applying big-data solution to a wide range of DoD applications including ISR, predictive logistics, decision support, and beyond.

% \begin{figure}[htbp!]
%     \centering
%     \includegraphics[width=0.48\textwidth]{figure/sdh.png}
%     \caption{SDH Program Concept and Structure \cite{SDH}}
%     \label{fig:sdh}
% \end{figure}

Hardware in the DARPA SDH program is expected to be configurable for executing different types of computations as efficiently as ASICs (under different configurations). Concomitantly, the software should be able to compile from high-level languages such as Python to generate executable code as well as the optimal hardware configuration. The main feature of this coupled software-hardware system is that it allows a workflow to change the hardware configuration at runtime. Further, the system is expected to be capable of  reusing the similar hardware configuration in executed workflows for new workflows. This requires the system to be able to store, analyze and retrieve historical logs on compilation and execution. The ultimate goal of the SDH program is to be able to take advantage of {\it data-dependent optimizations} that are even better than today's ASICs.

\subsection{Main Contributions}

To address the need for dynamic compiler optimizations of workflows, we propose a novel {\it dynamic and extensible  knowledge base} which 1) continuously gathers knowledge during execution of workflows 2) identifies {\it optimal} implementations of workflows on  {\it optimal} (available) hardware configurations, given the sequence of their constituting steps, based on compile-time and run-time queries.

% Starting with a list of workflows, we manually profile them and insert the identified knowledge into the knowledge base, which creates the initial knowledge base. At runtime, knowledge base keeps track of the real time performance and gathers inputs from other components of the compiler. 
%The correction on identified knowledge as well as the adding of unidentified knowledge update the knowledge base. 
Our knowledge base is dynamic and extensible, i.e., it keeps updating its contents at runtime so that the knowledge stored is more accurate, and it evolves with time.
% Finally, we will describe the implementation of the knowledge base using the C++ Boost Library and its performance in terms of query response time.
We conduct experiments showing that the query time of the knowledge base is linear with its size. Combined with the result that a small number of algorithmic building blocks (kernels) can cover most workflows, we conclude that our knowledge base does not need to be large, and so it can quickly answer queries from other components of the compiler. 

To summarize, our main contributions are as follows:

\begin{enumerate}
    \item We propose a rich labeled tripartite network representation of the knowledge base, that gathers the knowledge of optimized implementations of key algorithmic steps and kernels on various parameterized hardware.
    \item We design a query interface so that other components of the compiler can interact with the knowledge base at compile-time as well as execution-time.
    \item We present the implementation of the knowledge base using C++ Boost library.
    \item We demonstrate that the knowledge base is capable of answering queries within $1ms$ even with large number of workflows.
    \item We demonstrate that a small number of kernels can capture majority of workflows, suggesting that our knowledge base can be lightweight.
\end{enumerate}
\section{Background and Related Work}

The key to build the software defined hardware that can achieve both high programmability and efficiency is to build compilers for high-level programming language. Such compilers compile the programs before and during runtime. The compilers generates hardware configuration when the computation and communication patterns changes in the programs.

\subsection{DDARING Project}

\begin{figure}
    \centering
    \includegraphics[width=0.48\textwidth]{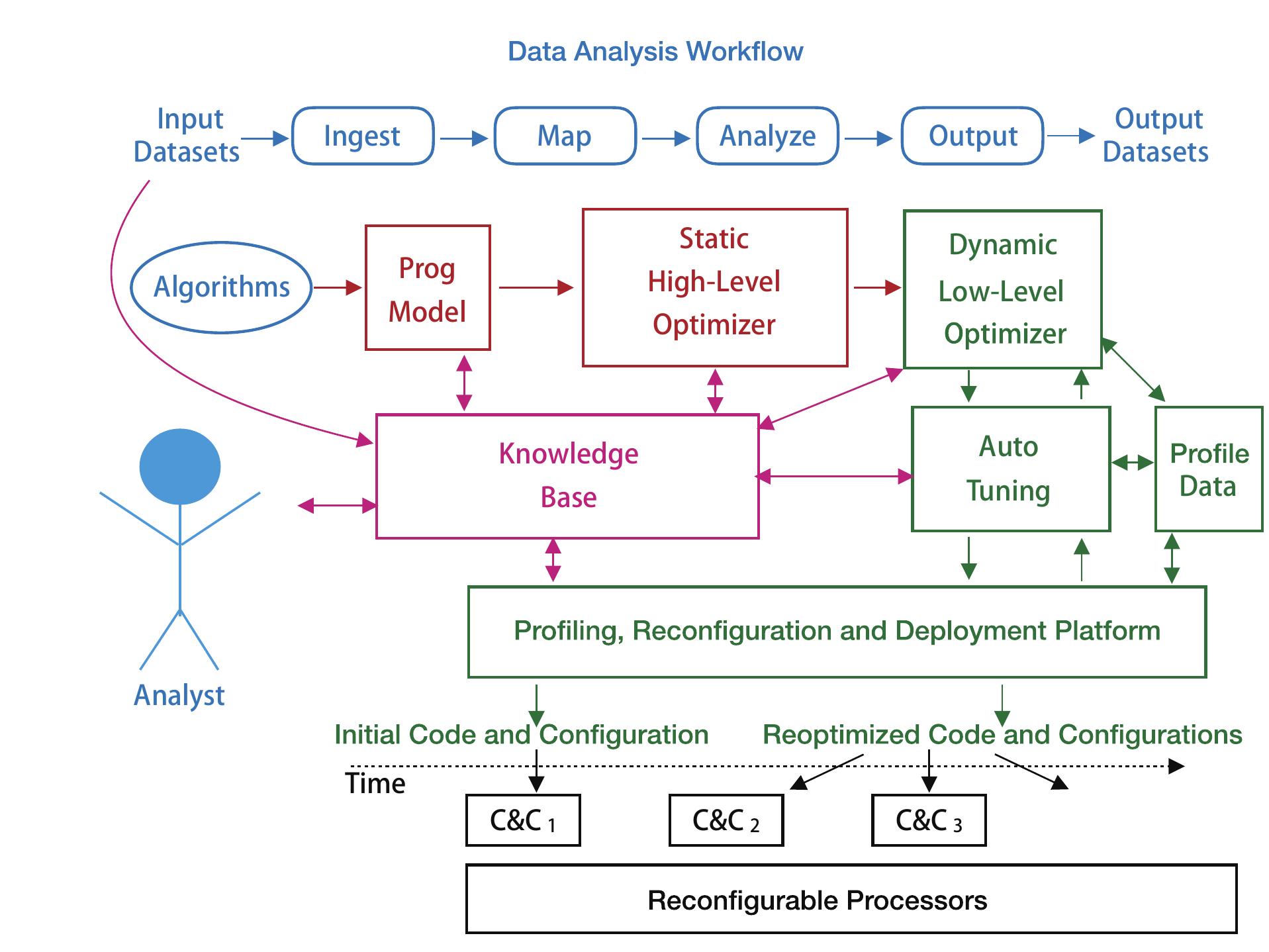}
    \caption{Overview of DDARING Project \cite{DDARING}}
    \label{fig:ddaring}
\end{figure}

One of the approaches for the dynamic hardware and software compilers of the SDH \cite{SDH} program is the DDARING project \cite{DDARING}. The knowledge base is designed as a center component in the DDARING project. Figure \ref{fig:ddaring} shows an overview of the technical approach of the DDARING project. The goal is to accelerate the workflows for data-intensive analyses to achieve near-ASIC performance. At the same time, the DDARING project can achieve the productivity that analysts have come to expect from modern high level problem-solving environments such as Julia and Python. The DDARING project encapsulates a unification of software compilation and hardware reconfiguration techniques, and is comprised of six components: the programming model, the knowledge base, the static high-level optimizer, the dynamic high-level optimizer, the auto-tuner and profiling, reconfiguration and deployment system. The workflows are first fitted into a programming model and compiled statically. At runtime, the dynamic optimizer changes the configuration of the hardware to achieve optimal compute efficiency. This process of static and dynamic compilation needs the supporting information from the metadata of the workflow and status on previous executions.

Specifically, developers would implement some workflows in a high-level programming language, i.e., Python. For the computational intensive parts that consume most execution time in the workflows, the developers instead implement these parts in the programming model built for DDARING called Intrepydd. There are also many pre-implemented high efficiency functions in Intrepydd language that the developers can use directly. Since most time of the workflows is spent in the computational intensive parts, this helps the developers to leverage the maximum convenience of high-level programming language. For performance, the Intrepydd code is first optimized by a static data-aware optimizer then optimized by a dynamic kernel reoptimizer and imported to the Python program. These two optimizers endow these computational intensive parts to leverage the advantages of reconfigurable hardware and to achieve near-ASIC performance. 

In the DDARING project, the key component is the knowledge base. To support the static data-aware optimizer and dynamic kernel reoptimizer, the knowledge base captures both the meta-data in the workflows and historical performance of similar workflows. To achieve high performance, these two optimizers compile and reconfigure the hardware at runtime when necessary. The knowledge base also treats the knowledge in a fine-grained level to uphold these fine-grained optimizers. One workflow is divided into many parts while similar parts in different workflows are stored and analyzed together in the knowledge base. At compilation and runtime, the knowledge base gathers information and answers queries from other components. This requires the knowledge base to respond rapidly even with a large amount of stored contents. To store fine-grained knowledge and to respond in real time are the two main challenges when designing and implementing the knowledge base.

\subsection{Related Work}

%In the field of high performance computing, the current trend is to use the heterogeneous architecture with specialized hardware working as accelerators for different parts of the computation. Within these heterogeneous architectures, reconfigurable hardware is a nimble design. The capability of changing hardware configurations at runtime endues its ability to execute computational intensive workflows efficiently. However, using reconfigurable hardware is more challenging than non-reconfigurable hardware. To make reconfigurable hardware efficient, we need compilers and programming languages that support joint optimization of hardware and software. Traditional compilers assume a fixed processor. They can generate code for that processor from input programs and iteratively refine the code they produce by analysis of a runtime trace. This process allows for data-dependent optimization but is limited by the available processing hardware.
Task scheduling in heterogeneous hardware systems has been widely researched \cite{doi:10.1002/cpe.1631,Yang92pyrros:static,765092,476197}. Since the introduction of reconfigurable hardware systems, there also has been considerable research done in the field of task scheduling and hardware mapping \cite{Ahmadinia:2004:TSH:1016568.1016582,1511977,1515787,Bondalapati2000LoopPA}. 
% The capability of changing hardware configurations at runtime endues its ability to execute computational intensive workflows efficiently.
Compilers for reconfigurable hardware need to co-optimize processor configuration and code using a iterative procedure. The problem is that this optimization space is much larger than what traditional compilers have to contend with. Hence, it is necessary to have a knowledge base that remembers the explored design space and give advice on further space.

Existing research on predicting the performance of a workflow tends to focus on accurate prediction before execution based on the matadata or its reference performance on other hardware configuration. For instance, \cite{6970672} proposed an accurate way to predict GPU performance based on reference performance and data access pattern on CPU. 
%This is a good start for compilers of reconfigurable hardware. 
However, once the workflows have been executed on GPU several times, the compiler should gather these real performance metrics and analyze these data to predict more accurately for the next runs. 
%Another constraint is that viewing the workflows as a whole is a coarse grain way to build the model. As section \ref{sec:sdh} mentioned, to achieve near-ASICs performance, the compiler needs to reconfigure the hardware rapidly at runtime. 
Existing compilers designed for heterogeneous architectures, such as HEFT \cite{heft}, MATEHa \cite{mateha}, etc. decompose the workflows into separate tasks and optimize the task graph execution. 

These approaches assume a prior knowledge of communication and computation models and task decomposition. On the contrary, the knowledge base differentiates from these approaches that it is more comprehensive. The knowledge base focuses not just on optimal mapping but also on semi-supervised extraction of knowledge from the workflows and development of the performance models. The knowledge base also captures the historical runtime performance data and updates based on the analyses from these data. This allows the knowledge base to evolve across executions of different workflows and to gain performance with the captured knowledge. Eventually, the knowledge base aids the compiler to achieve near-ASIC performance on software defined hardware.
\section{Design}

\begin{figure}
    \centering
    \includegraphics[width=0.48\textwidth]{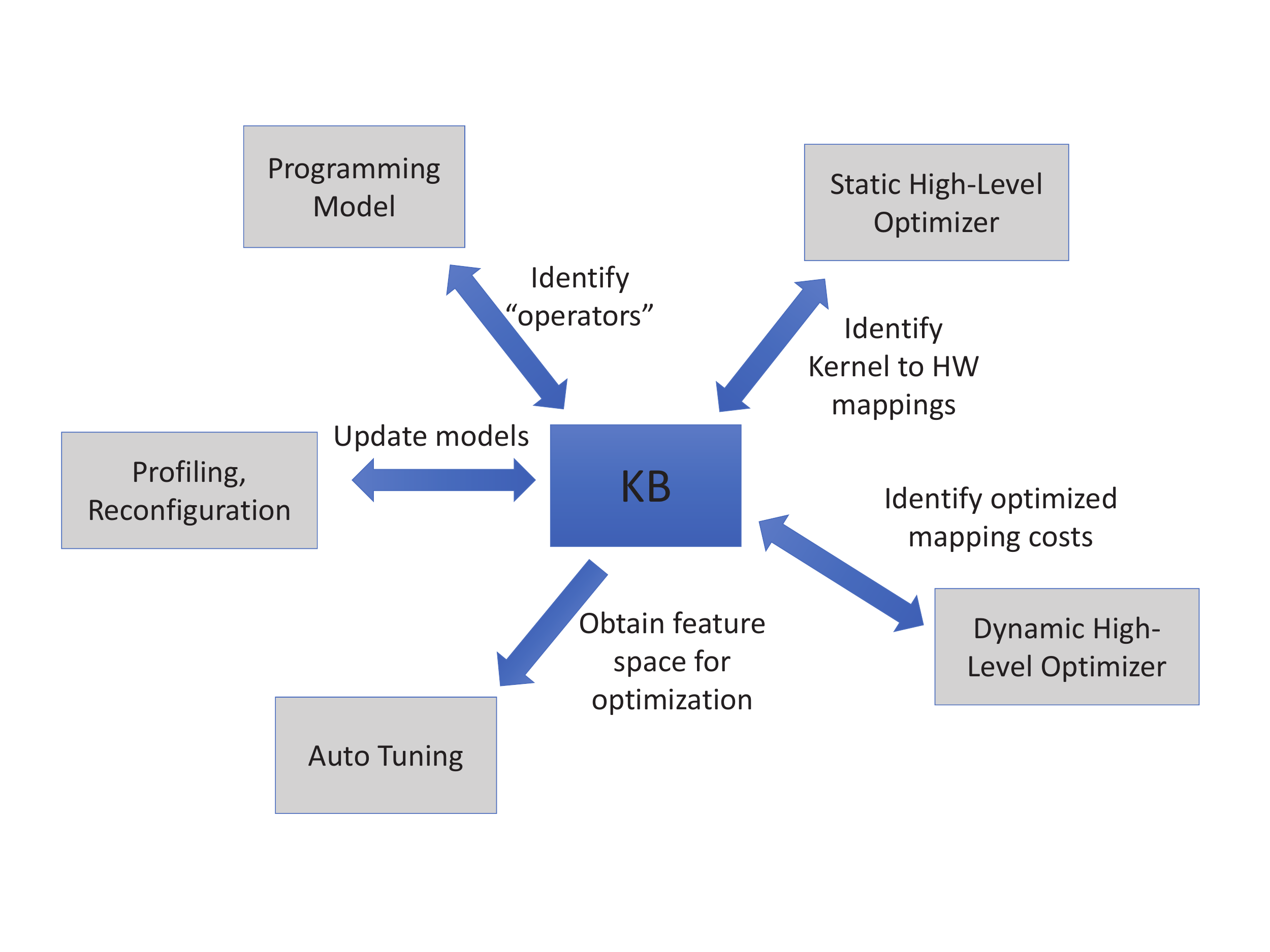}
    \caption{Interactions of the Knowledge Base with Other Components in the DDARING Project}
    \label{fig:KBinter}
\end{figure}

To construct the knowledge base, we view a workflow as a composition of several recurring time-consuming code segments which perform some specific tasks or operations. We denote such code segments as kernel primitives. The knowledge base stores extensive information about decomposition of distinct code modules (steps) into kernels and the optimal mapping of kernels to the underlying reconfigurable hardware.

\subsection{Representation}

\begin{figure}[htbp!]
    \centering
    \includegraphics[trim=1.5in 1in 1.5in 1in,clip, width=0.9\columnwidth]{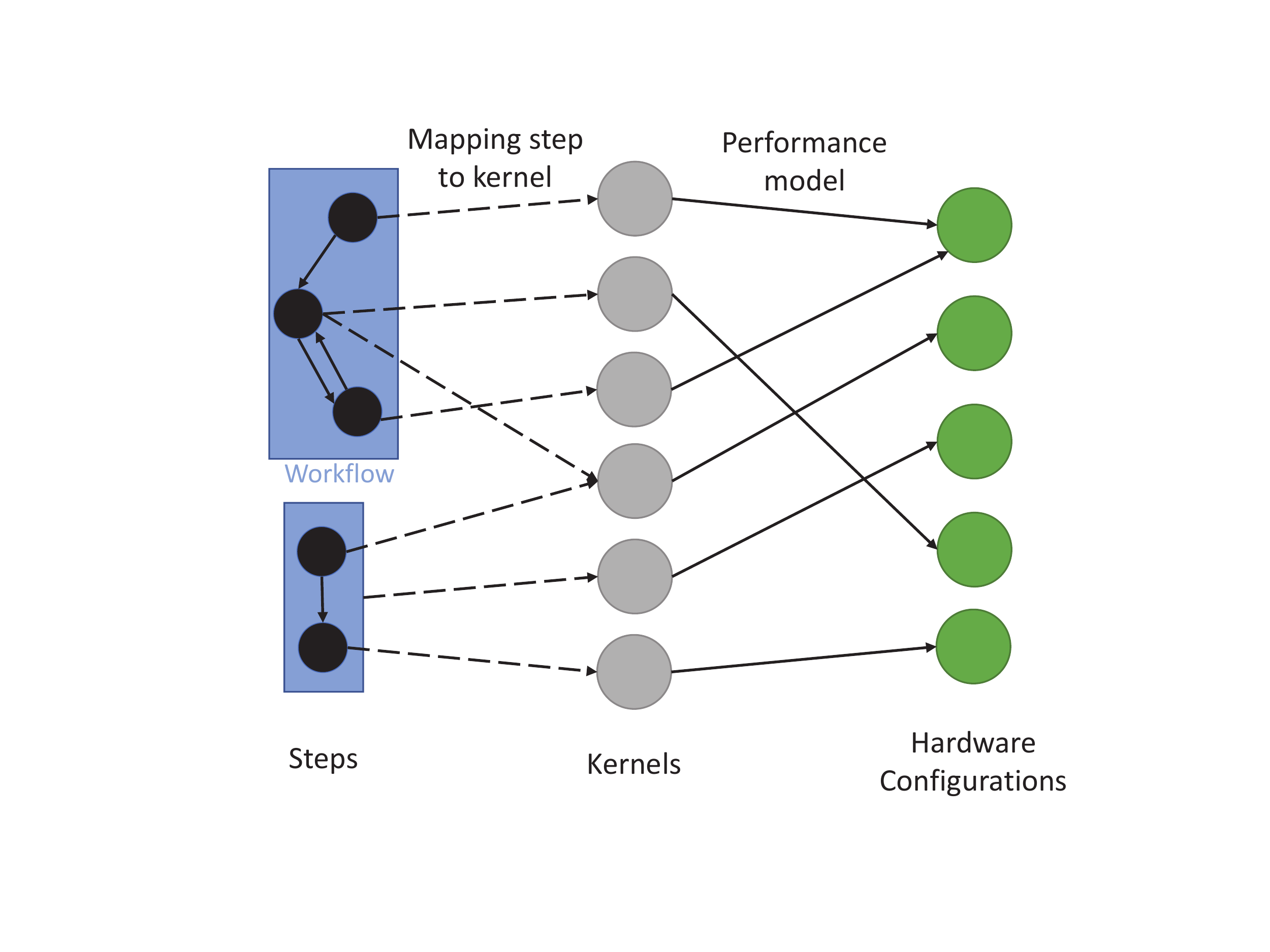}
    \caption{Tripartite Representation of Knowledge Base}
    \label{fig:kbtri}
\end{figure}

The knowledge base comprises of a rich tripartite graph representation $G(V_1,V_2,V_3,E)$ (TGR) as shown in figure \ref{fig:kbtri}. The first layer $V_1$ consists of Domain Specific Language level (DSL) steps. A step is a pattern of computation that can be mapped to one or more kernels. The set $V_1$ is the steps discovered in a codebase which includes a wide range of data intensive workflows. Each $u\in V_1$ is attributed with a feature vector $\vec{u_a}=(\{\mathcal{P}\},\{\mathcal{M}\})$ to quantitatively represent computational and data access patterns of the step (or part of the workflow) such as access pattern irregularity, data precision, computation to communication ratio and etc. A workflow is a finite-state-machine of the steps in $V_1$. A step may have multiple variants in terms of kernels.

Nodes in set $V_2$ represent bare bone tasks or kernels that form the core of the knowledge base. The connectivity between a step $u\in V_1$ and the set of kernels $t\in V_2$ delineates the lower level decomposition of $u$. Each edge $e=(u,t)$ will be attributed with a tuple $(w_e,d_{id})$. The weight assignment $w_e$ encodes meta-data of $u$ to be passed to $t$. The indexing function $d_{id}$ enables the proposed knowledge base representation to support more than one mapping of a step. For instance, a convolution operation can be represented as 1) a sequence of frequency domain transformation, multiplication and inverse transformation steps $(d_{id}=0)$, or 2) a series of sliding window dot products $(d_{id}=1)$. Frequently co-occurring kernels may be merged to form one kernel for optimized implementations.

The key characteristic of the TGR is that it exposes the resource requirements of the kernels that allow the knowledge base to obtain optimal program decomposition and task ordering across different segments of a given workflow.

The kernels $t\in V_2$ are computationally primitive units and feature  mappings to the underlying reconfigurable hardware platform and the building blocks in it. The hardware configurations and building blocks are represented as vertices $v\in V_3$. Note that a kernel can be mapped to one hardware configuration $(\mathcal T,\mathcal \mathcal C)$ in multiple ways $(\mathcal T)$ with different performance costs $(C)$. This is desirable as it would provide ways to schedule a task even if the optimal hardware block for execution is busy. 
% To support such scheduling, we insert an edge $e_0=(t,v)\quad\forall v|t$ can be executed on $v$ and label $e_0$ with performance and communication cost of the respective mapping.
Vertices in $V_3$ also store their cost of reconfiguration which is taken into account while computing the optimal mapping and overall cost of kernel execution.

\begin{figure}[htbp!]
    \centering
    \includegraphics[trim=1in 1in 0 1in,clip, width=\columnwidth]{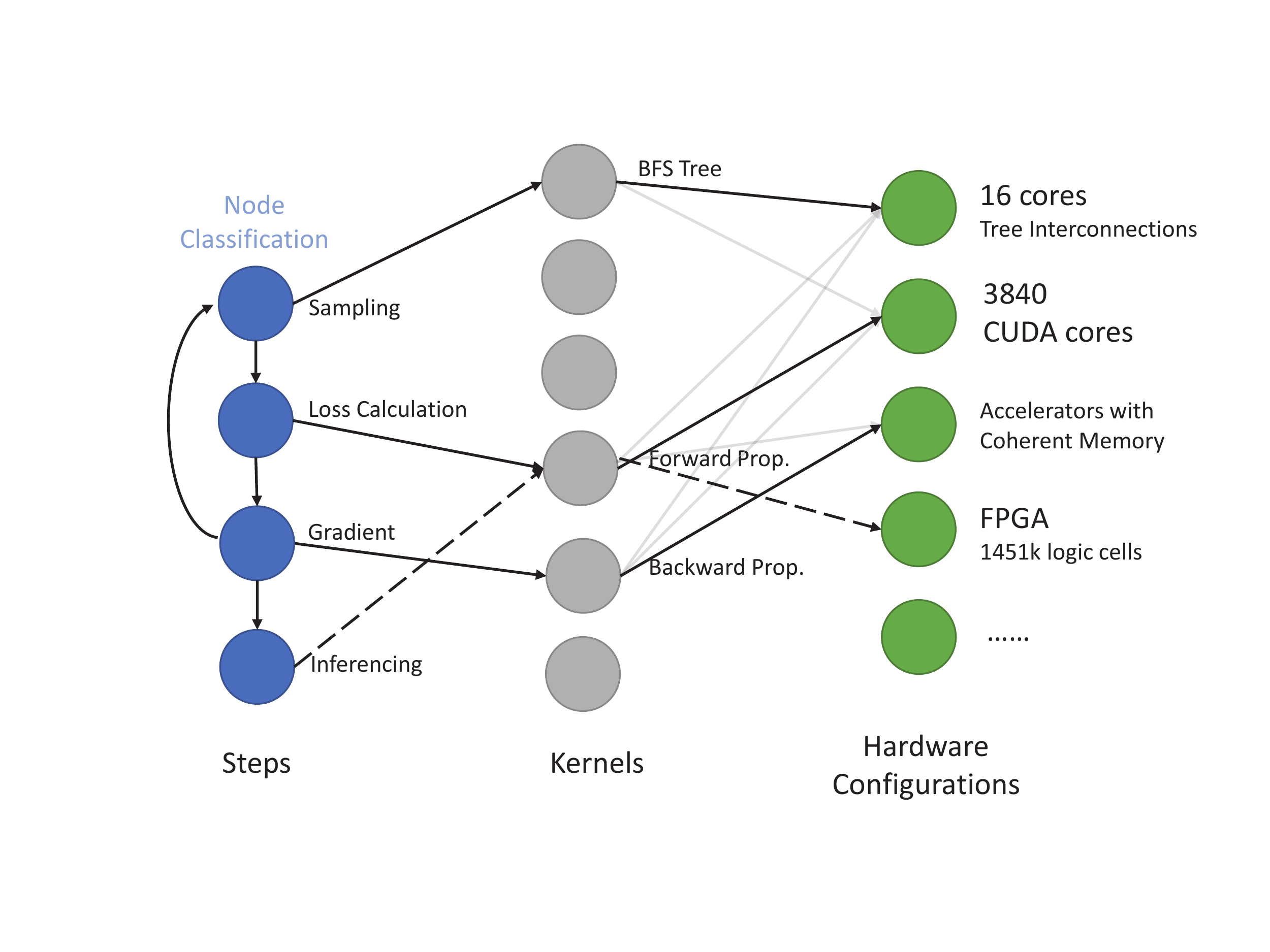}
    \caption{Running the Node Classification Workflow using the Knowledge Base}
    \label{fig:gsage}
\end{figure}

Figure \ref{fig:gsage} shows how the knowledge base stores the node classification workflow. This workflow is manually profiled and its knowledge is inserted into the knowledge base by human hands. As in the first layer $V_1$, the workflow is decomposed into four steps: sampling of the neighbors; evaluating the loss function; applying the gradients; and inferencing on test nodes. The first three steps are executed recursively until some termination criteria, followed by the execution of the last step. Based on the metadata of each steps, sampling is mapped to the BFS tree. Loss and gradient are each mapped to the corresponding neural network propagation. Note the dotted line from inference to forward propagation to the FPGA. Since inference does not required high precision as its metadata suggests, the knowledge base captures it and maps it to the FPGA hardware through some quantization technique to fit the FPGA resources.

\subsection{Creation}

The creation of knowledge base consists of two parts: offline discovery and dynamic updates. 

For offline discovery, we start with some chosen typical workflows. These workflows are first decomposed into steps and the steps are then mapped to all possible hardware configurations. We manually profile and construct the performance model for each pair and enter the model into the knowledge base. To begin the construction of the knowledge base, we have two types of kernel: (i) kernels with well-known optimized mappings to hardware such as dense matrix computation, sparse-matrix computation, FFT, etc. (ii) kernels for which optimal mappings are not known. We apply clustering algorithm to find associations of Type (ii) kernels with Type (i) kernels so that we can identify mapping of the input program to the kernels. Intuitively, if an unknown step $P$ has high similarities with a known step $M$, it is likely that the optimal hardware configuration for $M$ will achieve high performance on $P$ as well. However, instead of using the configuration of a single step, we use the similarities with multiple steps obtained by clustering, to generate performance cost estimates for kernels in $P$ for possible configurations $v\in V_3$. We also take into account the fact that a step is a sequence of kernels, and the ordering may result in different preferences of hardware configurations. To address such sequential nature of workflows that can be solved by using sequence labeling algorithms. In the SDH \cite{SDH} program, we profile the codebases provided by the HIVE \cite{HIVE} and D3M corpora and create the knowledge base accordingly.

For dynamic updates, the profiling and reconfiguration component in the DDARING project starts with the performance model obtained in offline discovery. As it analyzes and runs the program, it will update its own prediction model and also share the updates with the knowledge base dynamically. Thus, the knowledge base will not only be able to update existing cost models, but also create kernels and learn optimal configurations at runtime.

\subsection{Interaction with Other Components}
\label{sec:inter}

The knowledge base is the hub of the DDARING project. All other components provide inputs to, or consult queries from the knowledge base.  Figure \ref{fig:KBinter} shows the interactions between the knowledge base and the other five components. Specifically,

\begin{itemize}
    \item Programming Model: The knowledge base obtains input datasets and supervision from analyst and bootstrap. The knowledge base analyzes the programming model to specify and segments steps into kernels. 
    \item Static High-Level Optimizer: Static High-Level Optimizer consults the knowledge base to identify configurations for key steps. The knowledge base also uses the Static High-Level Optimizer in the offline discovery.
    \item Dynamic High-Level Optimizer: Dynamic High-Level Optimizer consults the knowledge base to obtain mappings of kernels to hardware building blocks and to construct reconfiguration cost models and to perform low-level optimizations.
    \item Auto Tuning:  Auto Tuning consults the knowledge base to enumerate the set of available optimizations and configurations to find optimal tuning of relevant parameters.
    \item Profiling, Reconfiguration: Profiling, Reconfiguration and Deployment Platform interacts with the knowledge base to obtain a previously learned performance model and update it online.
\end{itemize}

These input and output are expected to occur sequentially in each stage of the workflows. The interactions between programming model and static high-level optimizer happen before starting the execution of the workflow. The interaction between dynamic high-level optimizer happens before the execution of a kernel. The interaction between auto tuning happens at the runtime of a kernel. The interaction between profiling and reconfiguration happens after a kernel finishes execution. 
\section{Implementation}

The knowledge base is implemented using the Boost Graph Library \cite{Boost}, which provides generic and STL-like graph interface and graph components for C++. We choose to use C++ to implement the knowledge base since the code to execute on reconfigurable hardware is also C++. To ensure good extendibility both in contents and functionality, the knowledge base is stored as a collection of objects. We implement five C++ classes:

\begin{itemize}
    \item Step\_t: represents steps in a workflow. This class stores the static features of the step and supplies the features to the map from a step to a kernel.
    \item Kernel\_t: represents identified and unidentified kernels. 
    \item Hardware\_t: represents hardware configuration. This class stores hardware characteristics needed when predicting the performance model.
    \item Kernel\_map\_t: represents the map from step to kernel. This class stores the learnable models give advice on whether one kernel can be used to accomplish the step.
    \item Performance\_model\_t: represents the map from kernel to hardware configurations. This class stores the learnable models that predict the execution time and energy requirement of executing one kernel on the hardware configuration based on the metadata of the workflow. This model also give advice on which hardware configuration to use for a given kernel.
\end{itemize}

%Each entry in the knowledge base is stored by the corresponding subclass inherited from its main class. For example, a linear performance model would be a subclass of Performance\_model\_t. The unidentified (or identified at runtime) subtype would be stored in its parent class. 

To realize the tripartite representation of the knowledge base, we use adjacency list with customized vertex properties and edge properties. As discussed in section \ref{sec:inter}, we do not consider concurrent queries or updates to the knowledge base. Hence we do not add any locks to these classes or the tripartite graph. Each node in the graph represents either a step, a kernel or a hardware configuration. For node and edge properties, there are indicators distinguishing different classes of the nodes, as well as pointers that point to the subclass (or parent class) of the node. The definition of the Boost graph, vertex properties and node properties are shown in listing \ref{lst:code}. The graph representation and the classes can be saved and loaded to or from an XML file so that the identified knowledge can be stored and reused. This is helpful when the developers want to load some pre-learned knowledge base and update it after discovering new kernels and performance models. 

\begin{figure}
\begin{lstlisting}[caption={Code Snippet on Defining the Knowledge Base },label={lst:code}]
// one vertex captures a step, a kernel or a hardware
typedef struct vertex_properties
{
	bool is_step = false, is_kernel = false, is_hardware = false;
	int id;
	Kernel_t *kernel;
	Hardware_t *hardware;
	Step_t *step;
}vertex_properties_t;

// one edge captures a performance model or a kernel map
typedef struct edge_properties
{
	bool is_performance_model = false, is_kernel_map = false;
	int id;
	Performance_model_t *performance_model;
	Kernel_map_t *kernel_map;
}edge_properties_t;

typedef boost::adjacency_list<boost::vecS, boost::vecS, boost::undirectedS, vertex_properties_t, edge_properties_t> graph_t;
\end{lstlisting}
\end{figure}
\section{Performance}

The knowledge base should be able to answer queries from other components in the compiler in real-time. We evaluate the performance of the knowledge base using three types of queries,

\begin{itemize}
    \item Type 1 query: For a given step in the knowledge base, return all kernels linked with that step.
    \item Type 2 query: For a given kernel and a given hardware configuration in the knowledge base, return the performance model for executing the kernel on the hardware configuration.
    \item Type 3 query: For a given kernel in the knowledge base, return the performance models for executing the kernels on all hardware configurations.
\end{itemize}

\begin{figure}[htbp!]
    \centering
    \includegraphics[width=0.48\textwidth]{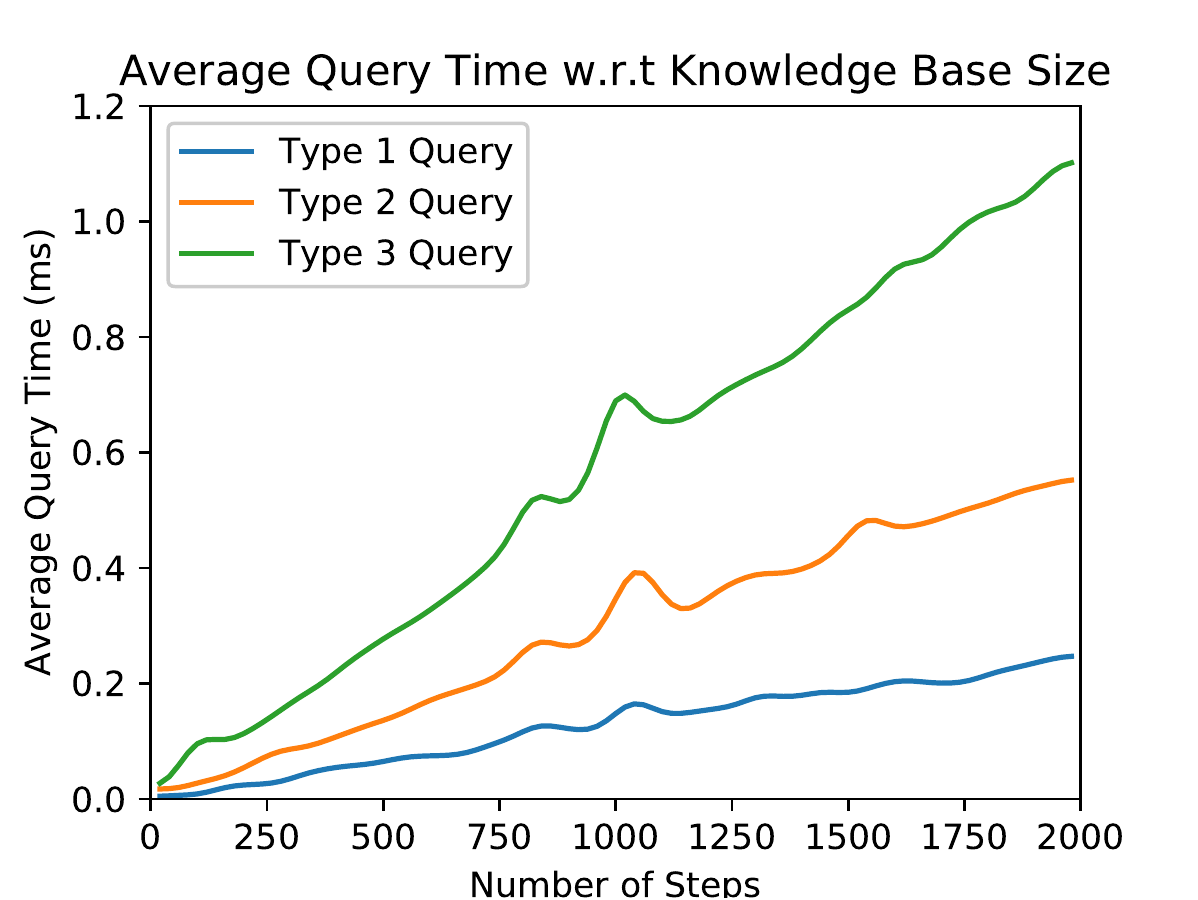}
    \caption{Average query time for different size of the knowledge base with probability $p=2/3$ and number of hardware configurations $h=10$.}
    \label{fig:time}
\end{figure}

We measure the average query time of these three types of queries in different sizes of knowledge bases. For each step, we assume that there are $k$ connections to $k$ kernels, where $k$ is chosen randomly from the geometric distribution with probability $p$. The average number $1/p$ represents the average number of kernels connected to one step. We fix the number of hardware configurations to be $h$ since the type of hardware is fixed in most scenario. Figure \ref{fig:time} shows the average query time for different sizes of the knowledge base. We can see that the query time increases linearly with the number of steps in the knowledge base. Type 3 query need the most time while type 2 query need the least time. All queries could be answered in $1ms$ even in a significantly large knowledge base (in reality, the number of steps in the current knowledge base is less than 200). The peak in average query time is due to the characteristics of cache. % the realization of the graph library itself.

We define the coverage of a workflow as the fraction of identified kernels in the knowledge base linked to all its decomposed steps. For example, a workflow is decomposed into two steps $S_1$ and $S_2$. $S_1$ is linked to kernel $K_1$ and $S_2$ is linked to kernel $K_2$ and $K_3$. If the knowledge base stores two kernels $K_1$ and $K_2$, we say that the coverage of this workflow in this knowledge base is $2/3$ and the uncoverage of this workflow is $1/3$. To model the coverage, we assume that when a randomly selected workflow wants to run a  kernel, the kernel is selected base on preferential attachment rule where the probability of selecting one kernel is proportional to the sum of frequency of this kernel and a fixed parameter $\lambda$.

\begin{figure}[htbp!]
    \centering
    \includegraphics[width=0.48\textwidth]{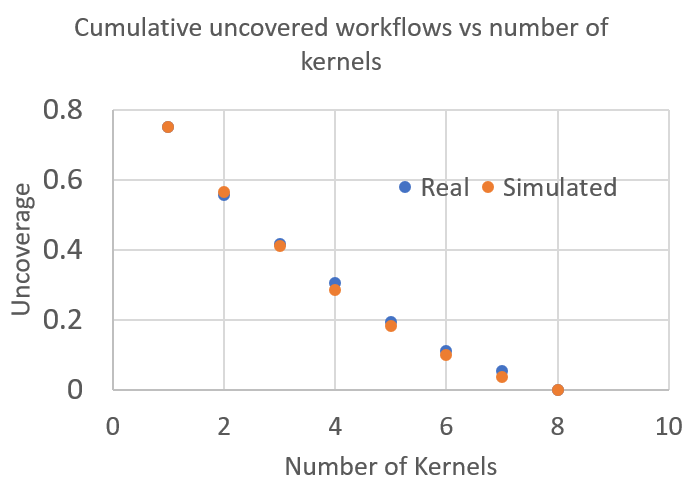}
    \caption{Uncoverage of SDH Workflows with Number of Kernels: Comparison of real values with the simulation-based predictions.}
    \label{fig:cover}
\end{figure}

Figure \ref{fig:cover} shows the predicted and real uncoverage of all workflows in the SDH \cite{SDH} codebases. The predicted uncoverage fits the real one accurately with $\lambda=2.5$. We can also see that the coverage increases exponentially with the number of kernels in the knowledge base, which means that a small knowledge base is enough to serve most of the workflows.

Although the average query time increases linearly with the size of the knowledge base, we do not need a large knowledge base when there is a large number of workflows. Combined with the fact that the query time of knowledge base is linear with the size of stored kernels, we can conclude that the knowledge base can answer queries rapidly regardless of the number of workflows captured in that knowledge base.

\section{Discussion and Conclusion}

We proposed a tripartite graph representation design and the C++ implementation of the knowledge base in the DDARING \cite{DDARING} project. The knowledge base takes input of datasets of codes, metadata, prediction model updates, and outputs mappings from steps to kernels and predictions of performance such as execution time, needed by other components in the DDARING project. 
As a part of the knowledge base, in a separate work, we have also developed light-weight neural-network based performance models to predict the execution times of kernels~\cite{naifeng}.
Our knowledge base is capable of responding to queries from other parts of the compiler within $1$ ms even for large number of kernels.
Although, we experimentally show that only using few kernels the knowledge base can cover a large number of workflows. 
In the next step of the DDARING project, we will show speedup results of executing workflows by the reconfigurable hardware and DDARING compiler using the knowledge base.

We believe that the knowledge base that gathers information on the metadata of the programs and learns mappings and performance models is central to compiler optimizations on reconfigurable devices. With the knowledge base, compilers also benefit when scheduling for non-reconfigurable devices such as the CPU and GPU. Even though the knowledge base is designed to support the DDARING project, our framework is generic and can be applied to most hardware platforms and programs.
\section*{Acknowledgement}
This  material  is  based  on  work  supported  by  the  Defense Advanced  Research  Projects  Agency  (DARPA)  under BAA Number HR001117S0055. Any opinions, findings and conclusions  or  recommendations  expressed  in  this  material  are those  of  the  authors  and  do  not  necessarily  reflect  the  views of DARPA.

\bibliographystyle{IEEEtran}
\bibliography{cite}

\end{document}